\documentclass[aps,prd,twocolumn,showkeys,amsmath,amssymb]{revtex4}
\usepackage{graphicx}
\usepackage{epstopdf}
\usepackage{multirow}
\usepackage[colorlinks,citecolor=blue,anchorcolor=red,menucolor=red,linkcolor=red,filecolor=red,runcolor=red,urlcolor=blue,frenchlinks=red]{hyperref}
\usepackage{soul}

\allowdisplaybreaks[3]

\def\f(s){\left[(\alpha+\beta)m_c^2-\alpha\beta s\right]}

\begin{document}
\title{Strong decays of fully-charm tetraquarks into di-charmonia}
%

\author{Hua-Xing Chen$^{1,2}$}
\author{Wei Chen$^3$}
\author{Xiang Liu$^{4,5}$}
\author{Shi-Lin Zhu$^{6,7,8}$\footnote{Corresponding author: zhusl@pku.edu.cn}}
\affiliation{
$^1$School of Physics, Southeast University, Nanjing 210094, China \\
$^2$School of Physics, Beihang University, Beijing 100191, China \\
$^3$School of Physics, Sun Yat-Sen University, Guangzhou 510275, China \\
$^4$School of Physical Science and Technology, Lanzhou University, Lanzhou 730000, China\\
$^5$Research Center for Hadron and CSR Physics, Lanzhou University and Institute of Modern Physics of CAS, Lanzhou 730000, China\\
$^6$School of Physics and State Key Laboratory of Nuclear Physics and Technology, Peking University, Beijing 100871, China \\
$^7$Collaborative Innovation Center of Quantum Matter, Beijing 100871, China \\
$^8$Center of High Energy Physics, Peking University, Beijing 100871, China}

\begin{abstract}
We study strong decays of the possible fully-charm tetraquarks recently observed by LHCb~\cite{Aaij:2020fnh}, and calculate their relative branching ratios through the Fierz rearrangement. Together with our previous QCD sum rule study~\cite{Chen:2016jxd}, our results suggest that the broad structure around $6.2$-$6.8$~GeV can be interpreted as an $S$-wave $cc\bar c \bar c$ tetraquark state with $J^{PC} = 0^{++}$ or $2^{++}$, and the narrow structure around 6.9~GeV can be interpreted as a $P$-wave one with $J^{PC} = 0^{-+}$ or $1^{-+}$. These structures were observed in the di-$J/\psi$ invariant mass spectrum, and we propose to confirm them in the di-$\eta_c$, $J/\psi h_c$, $\eta_c \chi_{c0}$, and $\eta_c \chi_{c1}$ channels. We also propose to search for their partner states having the negative charge-conjugation parity in the $J/\psi \eta_c$, $J/\psi \chi_{c0}$, $J/\psi \chi_{c1}$, and $\eta_c h_c$ channels.
\end{abstract}
\maketitle
\pagenumbering{arabic}

\section{Introduction}
\label{sec:intro}

Very recently, the LHCb Collaboration reported their results on possible fully-charm tetraquarks~\cite{Aaij:2020fnh}. They investigated the di-$J/\psi$ invariant mass spectrum, where they observed a broad structure ranging from 6.2 to 6.8~GeV and a narrow structure at around 6.9~GeV with a global significance of more than $5\sigma$. Especially, they assumed the latter narrow structure to be a resonance with the Breit-Wigner lineshape, and measured its mass and width to be either
\begin{eqnarray}
M &=& 6905 \pm 11 \pm 7 {\rm~MeV} \, ,
\\
\Gamma &=& 80 \pm 19 \pm 33 {\rm~MeV} \, ,
\end{eqnarray}
based on no-interference fit, or
\begin{eqnarray}
M &=& 6886 \pm 11 \pm 11 {\rm~MeV} \, ,
\\
\Gamma &=& 168 \pm 33 \pm 69 {\rm~MeV} \, ,
\end{eqnarray}
based on the simple model with interference.

\begin{table}
\begin{center}
\renewcommand{\arraystretch}{1.5}
\caption{Masses of $cc\bar c\bar c$ and $bb\bar b\bar b$ tetraquarks with various quantum numbers, calculated in Ref.~\cite{Chen:2016jxd} using the QCD sum rule method. Definitions of currents will be given in Sec.~\ref{sec:current}.
\label{tab:mass}}
\begin{tabular*}{8.4cm}{cccc}
\hline
~~~~$J^{PC}$ ~~~~& Currents &~~~~ $m_{c c\bar c \bar c}$ \mbox{(GeV)} ~~~~&~~~~ $m_{b b \bar b \bar b}$ \mbox{(GeV)} ~~~~
\\ \hline
$0^{++}$      & $J^{0^{++}}_1$              &  $6.44\pm0.15$              & $18.45\pm0.15$ \\
              & $J^{0^{++}}_2$              &  $6.46\pm0.16$              & $18.46\pm0.14$
\vspace{4pt} \\
$1^{+-}$      & $J^{1^{+-}}_{3\alpha}$      &  $6.51\pm0.15$              & $18.54\pm0.15$
\vspace{4pt} \\
$2^{++}$      & $J^{2^{++}}_{4\alpha\beta}$ &  $6.51\pm0.15$              & $18.53\pm0.15$
\vspace{4pt} \\
$0^{-+}$      & $J^{0^{-+}}_5$              &  $6.84\pm0.18$              & $18.77\pm0.18$ \\
              & $J^{0^{-+}}_6$              &  $6.85\pm0.18$              & $18.79\pm0.18$
\vspace{4pt} \\
$0^{--}$      & $J^{0^{--}}_7$              &  $6.84\pm0.18$              & $18.77\pm0.18$
\vspace{4pt} \\
$1^{-+}$      & $J^{1^{-+}}_{8\alpha}$      &  $6.84\pm0.18$              & $18.80\pm0.18$ \\
              & $J^{1^{-+}}_{9\alpha}$      &  $6.88\pm0.18$              & $18.83\pm0.18$
\vspace{4pt} \\
$1^{--}$      & $J^{1^{--}}_{10\alpha}$     &  $6.84\pm0.18$              & $18.77\pm0.18$ \\
              & $J^{1^{--}}_{11\alpha}$     &  $6.83\pm0.18$              & $18.77\pm0.16$ \\
\hline
\end{tabular*}
\end{center}
\end{table}

The above results are in a remarkable coincidence with our previous QCD sum rule predictions~\cite{Chen:2016jxd}, as partly shown in Table~\ref{tab:mass}. There we used eighteen diquark-antidiquark $\left([Q Q] [\bar Q \bar Q]\right)$ currents to perform QCD sum rule analyses, and calculated mass spectra of $cc\bar c\bar c$ and $bb\bar b\bar b$ tetraquark states. Although the currents used there do not explicitly contain derivatives, they can still have both positive- and negative-parities, and their quantum numbers can be $J^{PC} = 0^{++}/0^{-+}/0^{--}/1^{++}/1^{+-}/1^{-+}/1^{--}/2^{++}$.

There are altogether twelve currents of the positive parity, four of which well correspond to the $S$-wave $cc \bar c \bar c$ tetraquark states within the quark-model picture. We used them to perform QCD sum rule analyses, and the masses were predicted to be about 6.5~GeV~\cite{Chen:2016jxd}. Accordingly, the broad structure observed by LHCb at around $6.2$-$6.8$~GeV~\cite{Aaij:2020fnh} can be interpreted as an $S$-wave $cc\bar c \bar c$ tetraquark state. There are altogether seven currents of the negative parity. We also used them to perform QCD sum rule analyses, and the masses were predicted to be about 6.9~GeV~\cite{Chen:2016jxd}. Accordingly, the narrow structure observed by LHCb at around 6.9~GeV~\cite{Aaij:2020fnh} can be interpreted as a $P$-wave $cc\bar c \bar c$ tetraquark state.

Actually, the fully-heavy tetraquark states have been studied by some theorists since the 1980s~\cite{Chao:1980dv,Iwasaki:1975pv,Ader:1981db,Heller:1985cb,Badalian:1985es,Zouzou:1986qh,Lloyd:2003yc,Barnea:2006sd,Berezhnoy:2011xn}, but they did not receive much attention due to the absence of experimental data. However, with the running of LHC at 13 TeV and the Belle-II, searching for these exotic tetraquark states is becoming an important experimental issue. Besides the above LHCb experiment~\cite{Aaij:2020fnh}, in 2017 the CMS Collaboration had reported their measurement of an exotic structure in four lepton channel, and found an excess in $18.4 \pm 0.1 \pm 0.2$~GeV with a global significance of 3.6$\sigma$~\cite{Khachatryan:2016ydm,Yi:2018fxo}. This structure is a possible fully-bottom tetraquark state, but it was not confirmed in the latter LHCb and CMS experiments~\cite{Aaij:2018zrb,Sirunyan:2020txn}.

Driven by the above CMS and LHCb experiments~\cite{Khachatryan:2016ydm,Yi:2018fxo,Aaij:2018zrb,Sirunyan:2020txn,Aaij:2020fnh}, intensive theoretical studies have been performed to investigate fully-bottom and fully-charm tetraquark states~\cite{Chen:2016jxd,Karliner:2016zzc,Bai:2016int,Wang:2017jtz,Debastiani:2017msn,Anwar:2017toa,Esposito:2018cwh,Wu:2016vtq,Hughes:2017xie,Richard:2018yrm,Wang:2019rdo,Chen:2019dvd,Liu:2019zuc,Bedolla:2019zwg,Deng:2020iqw,liu:2020eha,Yang:2020rih,Wang:2020ols,Jin:2020jfc,Lu:2020cns}, while the conclusions are quite model dependent. For examples, the $b b \bar b \bar b$ tetraquark states are possible to lie below the di-bottomonium threshold according to Refs.~\cite{Chen:2016jxd,Karliner:2016zzc,Bai:2016int,Wang:2017jtz,Debastiani:2017msn,Anwar:2017toa,Esposito:2018cwh}, but they are not according to Refs.~\cite{Wu:2016vtq,Hughes:2017xie,Richard:2018yrm,Wang:2019rdo,Chen:2019dvd,Liu:2019zuc,Deng:2020iqw}. Due to these controversial issues, relevant experimental studies are crucial to understand them. From the theoretical side, studies on their decay properties are also useful and important~\cite{Becchi:2020mjz,Becchi:2020uvq}.

In this paper we shall study strong decay properties of fully-charm tetraquark states, based on our previous QCD sum rule study~\cite{Chen:2016jxd}. We shall investigate both $S$- and $P$-wave $c c \bar c \bar c$ tetraquark states. Our previous results in Ref.~\cite{Chen:2016jxd} suggest that their masses are above the di-$J/\psi$ and di-$\eta_c$ thresholds, so their strong decays into these two channels can happen, together with several other di-charmonia channels. Assuming them to be compact diquark-antidiquark $[c c] [\bar c \bar c]$ states, in the present study we shall apply the Fierz rearrangement of the Dirac and color indices to calculate their relative branching ratios. This method has been used in Refs.~\cite{Chen:2019wjd,Chen:2019eeq,Chen:2020pac} to study the $Z_c(3900)$, $X(3872)$, and $P_c$ states.

This paper is organized as follows. In Sec.~\ref{sec:current} we systematically construct diquark-antidiquark $\left([Q Q] [\bar Q \bar Q]\right)$ currents of both positive and negative parities, and use the Fierz rearrangement to transform them into meson-meson $\left([\bar Q Q] [\bar Q Q]\right)$ currents. Based on the obtained Fierz identities, in Sec.~\ref{sec:decay} we study their possible fall-apart decays, and calculate their relative branching ratios within the naive factorization scheme. The results are summarized in Sec.~\ref{sec:summary}.

\section{Currents and Fierz Identities}
\label{sec:current}

All the diquark-antidiquark $[Q Q] [\bar Q \bar Q]$ currents without derivatives have been systematically constructed in Ref.~\cite{Chen:2016jxd}, with only one current of $J^{PC} = 2^{++}$ missing:
\begin{eqnarray}
J_{\alpha\beta}^{2^{++}} &=& \Gamma_{\alpha\beta}^{\mu\nu}~Q_a^T C \sigma_{\mu\rho} Q_b~\bar Q_{a} \sigma_{\nu\rho} C \bar Q_{b}^T \, .
\end{eqnarray}
Here $Q_a$ is the heavy quark field with the color index $a$, and $\Gamma_{\alpha\beta}^{\mu\nu}$ is the projection operator,
\begin{eqnarray}
\Gamma^{\alpha\beta;\mu\nu} = g^{\alpha\mu} g^{\beta\nu} + g^{\alpha\nu} g^{\beta\mu} - {1\over2} g^{\alpha\beta} g^{\mu\nu}\, .
\end{eqnarray}

There are altogether twelve currents of the positive parity. Four of them well correspond to the $S$-wave $[QQ][\bar Q \bar Q]$ tetraquark states within the quark-model picture, and we shall only investigate these four currents in the present study. There are altogether seven currents of the negative parity, and we shall study all of them in the present study. Detailed expressions are given in the following subsections, together with the Fierz identities to transform them into meson-meson $\left([\bar Q Q] [\bar Q Q]\right)$ currents.

\subsection{Currents of the positive parity}

There are two $S$-wave diquarks, the ``good'' diquark of $J^P = 0^+$ and the ``bad'' one of $J^P = 1^+$ (other are ``worse'')~\cite{Jaffe:2004ph}. We can combine them to construct $S$-wave tetraquark states. To do this we follow the diquark-antidiquark model proposed in Refs.~\cite{Maiani:2004vq,Maiani:2014aja}. In this model the $S$-wave tetraquark states can be written in the spin basis as $|s, \bar s\rangle_J$, where $s = s_{QQ}$ and $\bar s = s_{\bar Q \bar Q}$ are the diquark and antidiquark spins, respectively.

There are altogether four $S$-wave $QQ\bar Q \bar Q$ tetraquark states, denoted as $|X;J^{PC}\rangle$:
\begin{eqnarray}
\nonumber
|X_1 ; 0^{++} \rangle &=& | 0, 0 \rangle_{0} \, ,
\\ |X_2 ; 0^{++} \rangle &=& | 1, 1 \rangle_0 \, ,
\label{eq:state1}
\\ \nonumber |X_3 ; 1^{+-} \rangle &=& | 1, 1 \rangle_1 \, ,
\\ \nonumber |X_4 ; 2^{++} \rangle &=& | 1, 1 \rangle_2 \, .
\end{eqnarray}

Similar to the quark-model picture, there are two $S$-wave diquark fields:
\begin{eqnarray}
&& Q_a^T C \gamma_5 Q_b \, ,~~~J^P = 0^+\, ,
\\ \nonumber && Q_a^T C \gamma_\mu Q_b \, ,~~~J^P = 1^+\, .
\end{eqnarray}
We can combine them to construct four tetraquark currents corresponding to Eqs.~(\ref{eq:state1}):
\begin{eqnarray}
J^{0^{++}}_1 &=& Q_a^T C \gamma_5 Q_b~\bar Q_{a} \gamma_5 C \bar Q_{b}^T \, ,
\label{def:current1}
\\
J^{0^{++}}_2 &=& Q_a^T C \gamma_\mu Q_b~\bar Q_{a} \gamma^\mu C \bar Q_{b}^T \, ,
\label{def:current2}
\\
J^{1^{+-}}_{3\alpha} &=& Q_a^T C \gamma^\mu Q_b~\bar Q_{a} \sigma_{\alpha\mu} \gamma_5 C \bar Q_{b}^T
\label{def:current3}
\\ \nonumber && ~~~~~~~~~~~~~~ - Q_a^T C \sigma_{\alpha\mu} \gamma_5 Q_b~\bar Q_{a} \gamma^\mu C \bar Q_{b}^T \, ,
\\
J^{2^{++}}_{4\alpha\beta} &=& \Gamma_{\alpha\beta}^{\mu\nu}~Q_a^T C \gamma_\mu Q_b~\bar Q_{a} \gamma_\nu C \bar Q_{b}^T \, .
\label{def:current4}
\end{eqnarray}
The tensor diquark field $Q_a^T C \sigma_{\mu\nu} \gamma_5 Q_b$ couples to both $J^P = 1^+$ and $1^-$ channels. However, its positive-parity component $Q_a^T C \sigma_{ij} \gamma_5 Q_b$ ($i, j=1, 2, 3$) gives the dominant contribution to $J^{1^{+-}}_{3\alpha}$. Hence, this current $J^{1^{+-}}_{3\alpha}$ corresponds to $|X_3 ; 1^{+-} \rangle = | 1, 1 \rangle_1$. Besides it, there exists another current directly corresponding to $|X_3 ; 1^{+-} \rangle$:
\begin{eqnarray}
J^{\prime1^{\pm-}}_{3\alpha\beta} &=& Q_a^T C \gamma_\alpha Q_b~\bar Q_{a} \gamma_\beta C \bar Q_{b}^T - \{ \alpha \leftrightarrow \beta \} \, ,
\end{eqnarray}
but this current contains both positive- and negative-parity components, so we do not investigate it in the present study. We refer to Ref.~\cite{Wang:2018poa} for its detailed discussions.

After applying the Fierz transformation on Eqs.~(\ref{def:current1}-\ref{def:current4}), we obtain:
\begin{eqnarray}
J^{0^{++}}_1 &=& -{1\over4} \eta_1^{0^{++}} -{1\over4} \eta_2^{0^{++}} -{1\over4} \eta_3^{0^{++}}
\label{eq:tran1}
\\ \nonumber && ~~~~~~~~~~~~~~~~~~ -{1\over4} \eta_4^{0^{++}} +{1\over8} \eta_5^{0^{++}} \, ,
\\
J^{0^{++}}_2 &=& \eta_1^{0^{++}} - \eta_2^{0^{++}} + {1\over2} \eta_3^{0^{++}} - {1\over2} \eta_4^{0^{++}} \, ,
\label{eq:tran2}
\\
J^{1^{+-}}_{3\alpha} &=& {3i} \eta_{6\alpha}^{1^{+-}} - \eta_{7\alpha}^{1^{+-}} \, ,
\label{eq:tran3}
\\
J^{2^{++}}_{4\alpha\beta} &=& {1\over2} \eta_{8\alpha\beta}^{2^{++}} - {1\over2} \eta_{9\alpha\beta}^{2^{++}} + {1\over2} \eta_{10\alpha\beta}^{2^{++}} \, .
\label{eq:tran4}
\end{eqnarray}
Some of these relations have been derived in Refs.~\cite{Chen:2006hy,Cui:2019roq}. Here $\eta_{1,2,3,4,5}^{0^{++}}$ are the meson-meson currents of $J^{PC}=0^{++}$, $\eta_{6\alpha,7\alpha}^{1^{+-}}$ are of $J^{PC}=1^{+-}$, and $\eta_{8\alpha\beta,9\alpha\beta,10\alpha\beta}^{2^{++}}$ are of $J^{PC}=2^{++}$:
\begin{eqnarray}
   \nonumber \eta_1^{0^{++}} &=& \bar Q_a Q_a~\bar Q_{b} Q_b \, ,
\\ \nonumber \eta_2^{0^{++}} &=& \bar Q_a \gamma_5 Q_a~\bar Q_{b} \gamma_5 Q_b \, ,
\\ \nonumber \eta_3^{0^{++}} &=& \bar Q_a \gamma_\mu Q_a~\bar Q_{b} \gamma^\mu Q_b \, ,
\\ \nonumber \eta_4^{0^{++}} &=& \bar Q_a \gamma_\mu \gamma_5 Q_a~\bar Q_{b} \gamma^\mu \gamma_5 Q_b \, ,
\\ \eta_5^{0^{++}} &=& \bar Q_a \sigma_{\mu\nu} Q_a~\bar Q_{b} \sigma^{\mu\nu} Q_b \, ,
\\ \nonumber \eta_{6\alpha}^{1^{+-}} &=& \bar Q_a \gamma_5 Q_a~\bar Q_{b} \gamma_\alpha Q_b \, ,
\\ \nonumber \eta_{7\alpha}^{1^{+-}} &=& \bar Q_a \gamma^\mu \gamma_5 Q_a~\bar Q_{b} \sigma_{\alpha\mu} Q_b \, ,
\\ \nonumber \eta_{8\alpha\beta}^{2^{++}} &=& \Gamma_{\alpha\beta}^{\mu\nu}~\bar Q_a \gamma_\mu \gamma_5 Q_a~\bar Q_{b} \gamma_\nu \gamma_5 Q_b \, ,
\\ \nonumber \eta_{9\alpha\beta}^{2^{++}} &=& \Gamma_{\alpha\beta}^{\mu\nu}~\bar Q_a \gamma_\mu Q_a~\bar Q_{b} \gamma_\nu Q_b \, ,
\\ \nonumber \eta_{10\alpha\beta}^{2^{++}} &=& \Gamma_{\alpha\beta}^{\mu\nu}~\bar Q_a \sigma_{\mu\rho} Q_a~\bar Q_{b} \sigma_{\nu\rho} Q_b \, .
\end{eqnarray}

Note that there are the same number of color-octet-color-octet meson-meson currents, while all of them can be related to the above color-singlet-color-singlet meson-meson currents, {\it e.g.},
\begin{equation}
\left(\begin{array}{c}
\eta_1^{\prime 0^{++}}
\\
\eta_2^{\prime 0^{++}}
\\
\eta_3^{\prime 0^{++}}
\\
\eta_4^{\prime 0^{++}}
\\
\eta_5^{\prime 0^{++}}
\end{array}\right)
=
\left(\begin{array}{ccccc}
-{7\over6} & -{1\over2} & -{1\over2} & {1\over2}  & -{1\over4}
\\
-{1\over2} & -{7\over6} & {1\over2}  & -{1\over2} & -{1\over4}
\\
-2         &  2         & {1\over3}  & 1          & 0
\\
2          &  -2        & 1          & {1\over3}  & 0
\\
-6         &  -6        & 0          & 0          & {1\over3}
\end{array}\right)
\left(\begin{array}{c}
\eta_1^{0^{++}}
\\
\eta_2^{0^{++}}
\\
\eta_3^{0^{++}}
\\
\eta_4^{0^{++}}
\\
\eta_5^{0^{++}}
\end{array}\right) \, ,
\end{equation}
where
\begin{eqnarray}
   \nonumber \eta_1^{\prime 0^{++}} &=& \bar Q_a \lambda^{ab} Q_b~\bar Q_c \lambda^{cd} Q_d \, ,
\\ \nonumber \eta_2^{\prime 0^{++}} &=& \bar Q_a \lambda^{ab} \gamma_5 Q_b~\bar Q_c \lambda^{cd} \gamma_5 Q_d \, ,
\\ \eta_3^{\prime 0^{++}} &=& \bar Q_a \lambda^{ab} \gamma_\mu Q_b~\bar Q_c \lambda^{cd} \gamma^\mu Q_d \, ,
\\ \nonumber \eta_4^{\prime 0^{++}} &=& \bar Q_a \lambda^{ab} \gamma_\mu \gamma_5 Q_b~\bar Q_c \lambda^{cd} \gamma^\mu \gamma_5 Q_d \, ,
\\ \nonumber \eta_5^{\prime 0^{++}} &=& \bar Q_a \lambda^{ab} \sigma_{\mu\nu} Q_b~\bar Q_c \lambda^{cd} \sigma^{\mu\nu} Q_d \, .
\end{eqnarray}

\subsection{Currents of the negative parity}

$P$-wave tetraquark states contain one unit of orbital excitation. In the diquark-antidiquark picture this orbital excitation can be between the diquark and antidiquark, or it can also be inside the diquark/antidiquark. Hence, $P$-wave tetraquark states are more complicated than $S$-wave ones. In a recent Ref.~\cite{liu:2020eha} the authors used a nonrelativistic potential quark model to systematically classify all the $P$-wave $QQ\bar Q \bar Q$ tetraquark states, where they found altogether twenty states. However, without using derivatives, we can not construct all their corresponding tetraquark currents. Instead, in the present study we shall use all the negative-parity tetraquark currents without derivatives to study their decay properties.

All the diquark-antidiquark $\left([Q Q] [\bar Q \bar Q]\right)$ currents of the negative parity have been systematically constructed in Ref.~\cite{Chen:2016jxd}. There are two currents of $J^{PC}=0^{-+}$:
\begin{eqnarray}
J^{0^{-+}}_5    &=& Q^T_aCQ_b ~ \bar{Q}_a \gamma_5 C \bar{Q}_b^T
\label{def:current5}
\\ \nonumber && ~~~~~~~~~~~~~~ + Q^T_aC\gamma_5Q_b ~ \bar{Q}_aC\bar{Q}_b^T \, ,
\\ J^{0^{-+}}_6 &=& Q^T_a C \sigma_{\mu\nu} Q_b ~ \bar{Q}_a \sigma^{\mu\nu} \gamma_5 C \bar{Q}^T_b \, .
\label{def:current6}
\end{eqnarray}
There is only one current of $J^{PC}=0^{--}$:
\begin{eqnarray}
J^{0^{--}}_7 &=& Q^T_a C Q_b ~ \bar{Q}_a\gamma_5C\bar{Q}_b^T
\label{def:current7}
\\ \nonumber && ~~~~~~~~~~~~~~ - Q^T_aC\gamma_5Q_b ~ \bar{Q}_aC\bar{Q}_b^T \, .
\end{eqnarray}
There are two currents of $J^{PC}=1^{-+}$:
\begin{eqnarray}
J^{1^{-+}}_{8\alpha} &=& Q^T_aC\gamma_\alpha\gamma_5 Q_b ~ \bar{Q}_a\gamma_5C\bar{Q}_b^T
\label{def:current8}
\\ \nonumber && ~~~~~~~~~~~~~~ + Q^T_aC\gamma_5Q_b ~ \bar{Q}_a\gamma_\alpha\gamma_5 C\bar{Q}_b^T \, ,
\\
J^{1^{-+}}_{9\alpha} &=& Q^T_aC\sigma_{\alpha\mu}Q_b ~ \bar{Q}_a\gamma^\mu C\bar{Q}^T_b
\label{def:current9}
\\ \nonumber && ~~~~~~~~~~~~~~ + Q^T_aC\gamma^\mu Q_b ~ \bar{Q}_a\sigma_{\alpha\mu}C\bar{Q}^T_b \, .
\end{eqnarray}
There are two currents of $J^{PC}=1^{--}$:
\begin{eqnarray}
J^{1^{--}}_{10\alpha} &=& Q^T_aC\gamma_\alpha\gamma_5 Q_b ~ \bar{Q}_a\gamma_5C\bar{Q}_b^T
\label{def:current10}
\\ \nonumber && ~~~~~~~~~~~~~~ - Q^T_aC\gamma_5Q_b ~ \bar{Q}_a\gamma_\alpha\gamma_5 C\bar{Q}_b^T \, ,
\\
J^{1^{--}}_{11\alpha} &=& Q^T_aC\sigma_{\alpha\mu}Q_b ~ \bar{Q}_a\gamma^\mu C\bar{Q}^T_b
\label{def:current11}
\\ \nonumber && ~~~~~~~~~~~~~~ - Q^T_aC\gamma^\mu Q_b ~ \bar{Q}_a\sigma_{\alpha\mu}C\bar{Q}^T_b \, .
\end{eqnarray}

Similar to Eqs.~(\ref{eq:state1}), we can write $P$-wave tetraquark states in the spin-parity basis as $|s^P, \bar s^{\bar P}\rangle_J$, where $P$ and $\bar P$ are the diquark and antidiquark parities, respectively. The above tetraquark currents correspond to the following $P$-wave $QQ\bar Q \bar Q$ tetraquark states:
\begin{eqnarray}
\nonumber |X_5 ; 0^{-+} \rangle &=& {1\over\sqrt2} \left( | 0^-, 0^+ \rangle_{0} + | 0^+, 0^- \rangle_{0} \right) \, ,
\\ \nonumber |X_6 ; 0^{-+} \rangle &=& | 1^{\pm}, 1^{\mp} \rangle_0 \, ,
\\ \nonumber |X_7 ; 0^{--} \rangle &=& {1\over\sqrt2} \left( | 0^-, 0^+ \rangle_{0} - | 0^+, 0^- \rangle_{0} \right) \, ,
\\ |X_8 ; 1^{-+} \rangle &=& {1\over\sqrt2} \left( | 1^-, 0^+ \rangle_{1} + | 0^+, 1^- \rangle_{1} \right) \, ,
\label{eq:state2}
\\ \nonumber |X_9 ; 1^{-+} \rangle &=& {1\over\sqrt2} \left( | 1^-, 1^+ \rangle_{1} + | 1^+, 1^- \rangle_{1} \right) \, ,
\\ \nonumber |X_{10} ; 1^{--} \rangle &=& {1\over\sqrt2} \left( | 1^-, 0^+ \rangle_{1} - | 0^+, 1^- \rangle_{1} \right) \, ,
\\ \nonumber |X_{11} ; 1^{--} \rangle &=& {1\over\sqrt2} \left( | 1^-, 1^+ \rangle_{1} - | 1^+, 1^- \rangle_{1} \right) \, .
\end{eqnarray}
The realistic physical states can be different from these states. However, if there exists some tetraquark current well (better) coupling to the physical state, the results extracted from this current should also be (more) consistent with that state. Hence, we investigate all the negative-parity tetraquark currents without derivatives in the present study, given that the internal structure of $P$-wave $QQ\bar Q \bar Q$ tetraquark states are still unknown.

After applying the Fierz transformation on Eqs.~(\ref{def:current5}-\ref{def:current11}), we obtain:
\begin{eqnarray}
J^{0^{-+}}_5 &=& -\eta_{11}^{0^{-+}} + {1\over4}\eta_{12}^{0^{-+}} \, ,
\label{eq:tran5}
\\
J^{0^{-+}}_6 &=& 6\eta_{11}^{0^{-+}} - {1\over2}\eta_{12}^{0^{-+}} \, ,
\label{eq:tran6}
\\
J^{0^{--}}_7 &=& - \eta_{13}^{0^{--}} \, ,
\label{eq:tran7}
\\
J^{1^{-+}}_{8\alpha} &=& - \eta_{14\alpha}^{1^{-+}} + i \eta_{15\alpha}^{1^{-+}} \, ,
\label{eq:tran8}
\\
J^{1^{-+}}_{9\alpha} &=& -3i \eta_{14\alpha}^{1^{-+}} + \eta_{15\alpha}^{1^{-+}} \, ,
\label{eq:tran9}
\\
J^{1^{--}}_{10\alpha} &=& \eta_{16\alpha}^{1^{--}} - i \eta_{17\alpha}^{1^{--}} \, ,
\label{eq:tran10}
\\
J^{1^{--}}_{11\alpha} &=& -3i \eta_{16\alpha}^{1^{--}} + \eta_{17\alpha}^{1^{--}} \, .
\label{eq:tran11}
\end{eqnarray}
Some of these relations have been derived in Refs.~\cite{Chen:2008ej,Chen:2008qw,Dong:2020okt}. Here $\eta_{11,12}^{0^{-+}}$ are the meson-meson currents of $J^{PC}=0^{-+}$, $\eta_{13}^{0^{--}}$ is of $J^{PC}=0^{--}$, $\eta_{14\alpha,15\alpha}^{1^{-+}}$ are of $J^{PC}=1^{-+}$, and $\eta_{16\alpha,17\alpha}^{1^{-+}}$ are of $J^{PC}=1^{--}$:
\begin{eqnarray}
\nonumber \eta_{11}^{0^{-+}} &=& \bar Q_a Q_a~\bar Q_{b} \gamma_5 Q_b \, ,
\\ \nonumber \eta_{12}^{0^{-+}} &=& \bar Q_a \sigma_{\mu\nu} Q_a~\bar Q_{b} \sigma^{\mu\nu} \gamma_5 Q_b \, ,
\\ \nonumber \eta_{13}^{0^{--}} &=& \bar Q_a \gamma_\mu Q_a~\bar Q_{b} \gamma^\mu \gamma_5 Q_b \, ,
\\ \eta_{14\alpha}^{1^{-+}} &=& \bar Q_a \gamma_5 Q_a~\bar Q_{b} \gamma_\alpha \gamma_5 Q_b \, ,
\\ \nonumber \eta_{15\alpha}^{1^{-+}} &=& \bar Q_a \gamma^\mu Q_a~\bar Q_{b} \sigma_{\alpha\mu} Q_b \, ,
\\ \nonumber \eta_{16\alpha}^{1^{--}} &=& \bar Q_a Q_a~\bar Q_{b} \gamma_\alpha Q_b \, ,
\\ \nonumber \eta_{17\alpha}^{1^{--}} &=& \bar Q_a \gamma^\mu \gamma_5 Q_a~\bar Q_{b} \sigma_{\alpha\mu} \gamma_5 Q_b \, .
\end{eqnarray}

\section{Relative Branching Ratios}
\label{sec:decay}

\begin{table*}[hbt]
\begin{center}
\renewcommand{\arraystretch}{1.5}
\caption{Couplings of charmonium operators to charmonium states. Color indices are omitted for simplicity.}
\begin{tabular}{ c  c  c  c  c  c}
\hline
~~~Operators~~~ & ~~~$J^{PC}$~~~ & ~~~Mesons~~~ & ~~~$J^{PC}$~~~ & ~~~Couplings~~~ & ~~~Decay Constants~~~
\\ \hline
$I^{S} = \bar c c$ & $0^{++}$ & $\chi_{c0}(1P)$  & $0^{++}$ &  $\langle 0 | I^S | \chi_{c0} \rangle = m_{\chi_{c0}} f_{\chi_{c0}}$  &  $f_{\chi_{c0}} = 343$~MeV~\mbox{\cite{Veliev:2010gb}}
\\[1mm]
$I^{P} = \bar c i\gamma_5 c$ & $0^{-+}$ &  $\eta_c$  & $0^{-+}$ &  $\langle 0 | I^{P} | \eta_c \rangle = \lambda_{\eta_c}$  &  $\lambda_{\eta_c} = {f_{\eta_c} m_{\eta_c}^2 \over 2 m_c}$
\\[1mm]
$I^{V}_\mu = \bar c \gamma_\mu c$ & $1^{--}$ &  $J/\psi$  & $1^{--}$ &  $\langle 0 | I^{V}_\mu | J/\psi \rangle = m_{J/\psi} f_{J/\psi} \epsilon_\mu$  &  $f_{J/\psi} = 418$~MeV~\mbox{\cite{Becirevic:2013bsa}}
\\[1mm]
$I^{A}_\mu = \bar c \gamma_\mu \gamma_5 c$ & $1^{++}$ &  $\eta_c$   & $0^{-+}$ &  $\langle 0 | I^{A}_\mu | \eta_c \rangle = i p_\mu f_{\eta_c}$  &  $f_{\eta_c} = 387$~MeV~\mbox{\cite{Becirevic:2013bsa}}
\\
                                                         &&  $\chi_{c1}(1P)$   & $1^{++}$ &  $\langle 0 | I^{A}_\mu | \chi_{c1} \rangle = m_{\chi_{c1}} f_{\chi_{c1}} \epsilon_\mu $  &  $f_{\chi_{c1}} = 335$~MeV~\mbox{\cite{Novikov:1977dq}}
\\[1mm]
$I^{T}_{\mu\nu} = \bar c \sigma_{\mu\nu} c$ & $1^{\pm-}$ &  $J/\psi$   & $1^{--}$   &  $\langle 0 | I^{T}_{\mu\nu} | J/\psi \rangle = i f^T_{J/\psi} (p_\mu\epsilon_\nu - p_\nu\epsilon_\mu) $  &  $f_{J/\psi}^T = 410$~MeV~\mbox{\cite{Becirevic:2013bsa}}
\\
                                                             &&  $h_c(1P)$   & $1^{+-}$ &  $\langle 0 | I^{T}_{\mu\nu} | h_c \rangle = i f^T_{h_c} \epsilon_{\mu\nu\alpha\beta} \epsilon^\alpha p^\beta $  &  $f_{h_c}^T = 235$~MeV~\mbox{\cite{Becirevic:2013bsa}}
\\ \hline
\end{tabular}
\label{tab:coupling}
\end{center}
\end{table*}

In this section we investigate possible decay channels of fully-charm tetraquark states, both qualitatively and quantitatively. According to the recent LHCb experiment~\cite{Aaij:2020fnh}, we assume the $S$- and $P$-wave $c c \bar c \bar c$ tetraquark states to have the masses about 6.5~GeV and 6.9~GeV, respectively.

\begin{figure}[hbt]
\begin{center}
\includegraphics[width=0.25\textwidth]{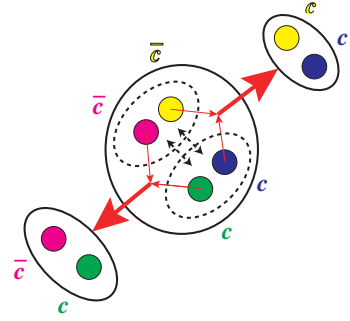}
\caption{The fall-apart decay of a compact diquark-antidiquark $[c c] [\bar c \bar c]$ state into two charmonium mesons. Quarks are shown in red/green/blue color, and antiquarks are shown in cyan/magenta/yellow color.}
\label{fig:diagram}
\end{center}
\end{figure}

As depicted in Fig.~\ref{fig:diagram}, when one heavy quark and one heavy antiquark meet each other and the rest heavy quark and antiquark also meet each other at the same time, a compact diquark-antidiquark $[cc] [\bar c \bar c]$ state can fall-apart decay into two charmonium mesons. This process can be described by the Fierz identities given in Eqs.~(\ref{eq:tran1}-\ref{eq:tran4}) and Eqs.~(\ref{eq:tran5}-\ref{eq:tran11}). To study it we need the couplings of charmonium operators to charmonium states, which have been well studied in the literature~\cite{pdg,Novikov:1977dq,Veliev:2010gb,Becirevic:2013bsa} and summarized here in Table~\ref{tab:coupling}.

%
%

Firstly, let us perform qualitative analyses. Take Eq.~(\ref{eq:tran1}) as an example, because both $\eta_3^{0^{++}}$ and $\eta_5^{0^{++}}$ can couple to the $J/\psi J/\psi$ channel, the current $J^{0^{++}}_1$ can also couple to this channel, so that the state $|X_1 ; 0^{++} \rangle$ can decay into the $J/\psi J/\psi$ final state. Similarly, we can derive six other possible channels to be $\eta_c \eta_c$, $\chi_{c0} \chi_{c0}$, $\chi_{c1} \chi_{c1}$, $h_c h_c$, $\eta_c \chi_{c1}$, and $J/\psi h_c$. Among them, the $J/\psi J/\psi$, $\eta_c \eta_c$, and $\eta_c \chi_{c1}$ channels are kinematically allowed.

In principle one needs the coupling of $J^{0^{++}}_1$ to $|X_1 ; 0^{++} \rangle$ as an input to quantitatively calculate partial decay widths of these channels. We define this coupling to be
\begin{equation}
\langle 0 | J^{0^{++}}_1 | X_1 ; 0^{++} \rangle = f_{X_1} \, ,
\end{equation}
while it is not necessary any more if one only calculates relative branching ratios. Moreover, because couplings of meson operators to meson states are well studied but couplings of tetraquark currents to tetraquark states are not, the decay constant $f_{X_1}$ is not so well determined compared to the meson decay constants listed in Table~\ref{tab:coupling}. Accordingly, relative branching ratios can be calculated more reliably than partial decay widths.

To calculate relative branching ratios, we just need to keep $f_{X_1}$ as an unfixed parameter, calculate partial decay widths, and finally remove $f_{X_1}$. Still take Eq.~(\ref{eq:tran1}) as an example, the couplings of $|X_1 ; 0^{++} \rangle$ to the $J/\psi J/\psi$ and $\eta_c \eta_c$ channels can be extracted from it to be:
\begin{eqnarray}
&& \langle X_1(p); 0^{++} | J/\psi(p_1,\epsilon_1)~J/\psi(p_2,\epsilon_2) \rangle
\\ \nonumber &\propto& f_{X_1} \times \epsilon_1^\mu \epsilon_2^\nu ~ \Big( - {1\over2} m_{J/\psi}^2 f_{J/\psi}^2 g_{\mu\nu}
\\ \nonumber && ~~~~~ - {1\over2} (f_{J/\psi}^{T})^2 p_1 \cdot p_2 g_{\mu\nu} + {1\over2} (f_{J/\psi}^{T})^2 p_{1\nu} p_{2\mu} \Big) \, ,
\\
&& \langle X_1(p); 0^{++} | \eta_c(p_1)~\eta_c(p_2) \rangle
\\ \nonumber &\propto& f_{X_1} \times \Big( {1\over2} \lambda_{\eta_c}^2 + {1\over2} f_{\eta_c}^2 p_1 \cdot p_2 \Big) \, .
\end{eqnarray}
After calculating the two partial decay widths $\Gamma_{|X_1 ; 0^{++} \rangle \to J/\psi J/\psi}$ and $\Gamma_{|X_1 ; 0^{++} \rangle \to \eta_c \eta_c}$, we can remove the parameter $f_{X_1}$ and obtain:
\begin{equation}
{\mathcal{B}(| X_1 ; 0^{++} \rangle \rightarrow J/\psi J/\psi) \over \mathcal{B}(| X_1 ; 0^{++} \rangle \rightarrow \eta_c \eta_c)} = 1 : 0.45 \, .
\end{equation}
Similarly, we can add the $\eta_c \chi_{c1}$ channel and obtain:
\begin{eqnarray}
\mathcal{B}(| X_1 ; 0^{++} \rangle &\rightarrow& J/\psi J/\psi ~:~ \eta_c \eta_c ~:~ \eta_c \chi_{c1}\,)
\label{eq:X1}
\\ \nonumber &=& ~~~~~1~~~~~ : ~\,0.45~ : ~2.1 \times 10^{-5}  \, .
\end{eqnarray}
Following the same procedures, we shall separately investigate the $S$- and $P$-wave $c c \bar c \bar c$ tetraquark states in the following subsections.

\subsection{$S$-wave $c c \bar c \bar c$ states}

Firstly, we use the Fierz identities given in Eqs.~(\ref{eq:tran2}-\ref{eq:tran4}) to perform qualitative analyses:
\begin{itemize}

\item Eq.~(\ref{eq:tran2}) suggests the possible decay channels of $|X_2 ; 0^{++} \rangle$ to be $J/\psi J/\psi$, $\eta_c \eta_c$, $\chi_{c0} \chi_{c0}$, $\chi_{c1} \chi_{c1}$, and $\eta_c \chi_{c1}$. Among them, the $J/\psi J/\psi$, $\eta_c \eta_c$, and $\eta_c \chi_{c1}$ channels are kinematically allowed.

\item Eq.~(\ref{eq:tran3}) suggests the possible decay channels of $|X_3 ; 1^{+-} \rangle$ to be $J/\psi \eta_c$, $J/\psi \chi_{c1}$, $\eta_c h_c$, and $h_c \chi_{c1}$. Among them, only the $J/\psi \eta_c$ channel is kinematically allowed.

\item Eq.~(\ref{eq:tran4}) suggests the possible decay channels of $|X_4 ; 2^{++} \rangle$ to be $J/\psi J/\psi$, $\eta_c \eta_c$, $\chi_{c1} \chi_{c1}$, $h_c h_c$, $\eta_c \chi_{c1}$, and $J/\psi h_c$. Among them, the $J/\psi J/\psi$, $\eta_c \eta_c$, and $\eta_c \chi_{c1}$ channels are kinematically allowed.

\end{itemize}

Quantitatively, we assume masses of the $S$-wave $c c \bar c \bar c$ tetraquark states to be about 6.5~GeV, and obtain:
\begin{eqnarray}
\mathcal{B}(| X_2 ; 0^{++} \rangle &\rightarrow& J/\psi J/\psi ~:~ \eta_c \eta_c ~:~ \eta_c \chi_{c1}\,)
\label{eq:X2}
\\ \nonumber &=& ~~~~~1~~~~~ : ~~4.1~~ : ~8.6 \times 10^{-5}  \, ,
\\ \mathcal{B}(| X_4 ; 2^{++} \rangle &\rightarrow& J/\psi J/\psi ~:~ \eta_c \eta_c ~:~ \eta_c \chi_{c1}\,)
\label{eq:X4}
\\ \nonumber &=& ~~~~~1\,~~~~ : ~0.036~ : ~6.0 \times 10^{-4}  \, .
\end{eqnarray}
Relative branching ratios of $|X_3 ; 1^{+-} \rangle$ are not given, because we only derive one of its possible decay channels.

\subsection{$P$-wave $c c \bar c \bar c$ states}

Firstly, we use the Fierz identities given in Eqs.~(\ref{eq:tran5}-\ref{eq:tran11}) to perform qualitative analyses:
\begin{itemize}

\item Eqs.~(\ref{eq:tran5}) and (\ref{eq:tran6}) suggest the possible decay channels of $|X_5 ; 0^{-+} \rangle$ and $|X_6 ; 0^{-+} \rangle$ to be $J/\psi J/\psi$, $h_c h_c$, $\eta_c \chi_{c0}$, and $J/\psi h_c$. Among them, the $J/\psi J/\psi$, $\eta_c \chi_{c0}$, and $J/\psi h_c$ channels are kinematically allowed.

\item Eq.~(\ref{eq:tran7}) suggests the possible decay channels of $|X_7 ; 0^{--} \rangle$ to be $J/\psi \eta_c$ and $J/\psi \chi_{c1}$. These two channels are both kinematically allowed.

\item Eqs.~(\ref{eq:tran8}) and (\ref{eq:tran9}) suggest the possible decay channels of $|X_8 ; 1^{-+} \rangle$ and $|X_9 ; 1^{-+} \rangle$ to be $J/\psi J/\psi$, $\eta_c \eta_c$, $\eta_c \chi_{c1}$, and $J/\psi h_c$. All these channels are kinematically allowed.

\item Eqs.~(\ref{eq:tran10}) and (\ref{eq:tran11}) suggest the possible decay channels of $|X_{10} ; 1^{--} \rangle$ and $|X_{11} ; 1^{--} \rangle$ to be $J/\psi \eta_c$, $J/\psi \chi_{c0}$, $J/\psi \chi_{c1}$, $\eta_c h_c$, and $h_c \chi_{c1}$. Among them, the $J/\psi \eta_c$, $J/\psi \chi_{c0}$, $J/\psi \chi_{c1}$, and $\eta_c h_c$ channels are kinematically allowed.

\end{itemize}

Quantitatively, we assume masses of the $P$-wave $c c \bar c \bar c$ tetraquark states to be about 6.9~GeV, and obtain:
\begin{eqnarray}
\nonumber \mathcal{B}(|X_5 ; 0^{-+} \rangle &\rightarrow& J/\psi J/\psi ~:~ \eta_c \chi_{c0} ~:~ J/\psi h_c\,)
\\ &=& ~~~~~1~~~~~ : ~~0.69~~ : ~0.21  \, ,
\label{eq:X5}
\\ \nonumber \mathcal{B}(|X_6 ; 0^{-+} \rangle &\rightarrow& J/\psi J/\psi ~:~ \eta_c \chi_{c0} ~:~ J/\psi h_c\,)
\\ &=& ~~~~~1~~~~~ : ~~~6.2\,~~ : ~0.21  \, ,
\label{eq:X6}
\\ \nonumber \mathcal{B}(|X_7 ; 0^{--} \rangle &\rightarrow& J/\psi \eta_c ~:~ J/\psi \chi_{c1}\,)
\\ &=& ~~~~1\,~~~ : ~~~1.4 \, ,
\label{eq:X7}
\\ \nonumber \mathcal{B}(|X_8 ; 1^{-+} \rangle &\rightarrow& J/\psi J/\psi ~:~ \eta_c \eta_c ~:~ \eta_c \chi_{c1} ~:~ J/\psi h_c\,)
\\ \nonumber &=& ~~~~~\,1~~~~~ : ~0.11~ : ~~0.36\,~ : ~0.30 \, ,
\\ \label{eq:X8}
\\ \nonumber \mathcal{B}(|X_9 ; 1^{-+} \rangle &\rightarrow& J/\psi J/\psi ~:~ \eta_c \eta_c ~:~ \eta_c \chi_{c1} ~:~ J/\psi h_c\,)
\\ \nonumber &=& ~~~~~1~~~~~ : ~~1.0~~ : ~~\,3.2\,~~ : ~0.30 \, ,
\\ \label{eq:X9}
\\ \nonumber \mathcal{B}(|X_{10} ; 1^{--} \rangle &\rightarrow& J/\psi \eta_c ~:~ J/\psi \chi_{c0} ~:~ J/\psi \chi_{c1} ~:~ \eta_c h_c\,)
\\ \nonumber &=& ~~~~1\,~~~ : ~~~\,0.79~~~ : ~~~~1.5\,~~~ : ~0.43 \, ,
\\ \label{eq:X10}
\\ \nonumber \mathcal{B}(|X_{11} ; 1^{--} \rangle &\rightarrow& J/\psi \eta_c ~:~ J/\psi \chi_{c0} ~:~ J/\psi \chi_{c1} ~:~ \eta_c h_c\,)
\\ \nonumber &=& ~~~~1\,~~~ : ~~~~7.1~~~~ : ~~~~1.5\,~~~ : ~0.43 \, .
\\ \label{eq:X11}
\end{eqnarray}

\section{Summary and discussions}
\label{sec:summary}

\begin{table*}[hbt]
\begin{center}
\renewcommand{\arraystretch}{1.6}
\caption{Relative branching ratios of the $S$- and $P$-wave $c c \bar c \bar c$ tetraquark states. In the second column we write these state in the spin-parity basis as $|s_{cc}^P, \bar s_{\bar c \bar c}^{\bar P} \rangle_J$; $s^P_{cc}$ and $\bar s^{\bar P}_{\bar c \bar c}$ are the diquark and antidiquark spins-parities, respectively; the subscripts $a/b$ denote $| X_{cc}, Y_{\bar c \bar c} \rangle_{J}^a = {1\over\sqrt2}\left( | X_{cc}, Y_{\bar c \bar c} \rangle_J + | Y_{cc}, X_{\bar c \bar c} \rangle_J \right)$ and $| X_{cc}, Y_{\bar c \bar c} \rangle_{J}^b = {1\over\sqrt2}\left( | X_{cc}, Y_{\bar c \bar c} \rangle_J - | Y_{cc}, X_{\bar c \bar c} \rangle_J \right)$. In the 3rd-7th columns we show branching ratios relative to the $J/\psi J/\psi$ channel, such as ${\mathcal{B}(X \to \eta_c \eta_c)\over\mathcal{B}(X \to J/\psi J/\psi)}$ in the 4th column. In the 8th-11th columns we show branching ratios relative to the $J/\psi \eta_c$ channel, such as ${\mathcal{B}(X \to J/\psi \chi_{c0})\over\mathcal{B}(X \to J/\psi \eta_c)}$ in the 9th column.}
\begin{tabular}{ c  c  c  c  c  c  c  c  c  c  c }
\hline
 && \multicolumn{9}{c}{Decay Channels}
\\ \cline{3-11}
~~$J^{PC}$~~ & ~~~Configuration~~~ & ~$J/\psi J/\psi$~ & ~~~$\eta_c \eta_c$~~~ & ~~$J/\psi h_c$~~ & ~~$\eta_c \chi_{c0}$~~ & ~~~~$\eta_c \chi_{c1}$~~~~ & ~~$J/\psi \eta_c$~~ & ~$J/\psi \chi_{c0}$~ & ~$J/\psi \chi_{c1}$~ & ~~~$\eta_c h_c$~~~
\\ \hline
$0^{++}$ & $X_1 = | 0^+_{cc}, 0^+_{\bar c \bar c} \rangle_0$ & $1$  & $0.45$ & -- & -- & $2.1 \times 10^{-5}$ & -- & -- & -- & --
\\
                          & $X_2 = | 1^+_{cc}, 1^+_{\bar c \bar c} \rangle_0$ & $1$  & $4.1$ & -- & -- & $8.6 \times 10^{-5}$ & -- & -- & -- & --
\\[1mm]
$1^{+-}$ & $X_3 = | 1^+_{cc}, 1^+_{\bar c \bar c} \rangle_1$ & -- & -- & -- & -- & -- & $1$ & -- & -- & --
\\[1mm]
$2^{++}$ & $X_4 = | 1^+_{cc}, 1^+_{\bar c \bar c} \rangle_2$ & $1$  & $0.036$ & -- & -- & $6.0 \times 10^{-4}$ & -- & -- & -- & --
\\[1mm]
$0^{-+}$ & $X_5 = | 0^-_{cc}, 0^+_{\bar c \bar c} \rangle_{0}^{a}$ & $1$  & -- &  $0.21$  & $0.69$ & -- & -- & -- & -- & --
\\
                          & $X_6 = | 1^\pm_{cc}, 1^\mp_{\bar c \bar c} \rangle_{0}$ & $1$  & -- &  $0.21$  & $6.2$ & -- & -- & -- & -- & --
\\[1mm]
$0^{--}$ & $X_7 = | 0^-_{cc}, 0^+_{\bar c \bar c} \rangle_{0}^{b}$ & -- & -- & -- & -- & -- & $1$ & -- & $1.4$ & --
\\[1mm]
$1^{-+}$ & $X_8 = | 1^-_{cc}, 0^+_{\bar c \bar c} \rangle_{1}^{a}$ & $1$  & $0.11$ &  $0.30$  & -- & $0.36$ & -- & -- & -- & --
\\
                          & $X_9 = | 1^-_{cc}, 1^+_{\bar c \bar c} \rangle_{1}^{a}$ & $1$  & $1.0$ &  $0.30$  & -- & $3.2$ & -- & -- & -- & --
\\[1mm]
$1^{--}$ & $X_{10} = | 1^-_{cc}, 0^+_{\bar c \bar c} \rangle_{1}^{b}$ & -- & -- & -- & -- & -- & $1$ & $0.79$ & $1.5$ & $0.43$
\\
                          & $X_{11} = | 1^-_{cc}, 1^+_{\bar c \bar c} \rangle_{1}^{b}$ & -- & -- & -- & -- & -- & $1$ & $7.1$ & $1.5$ & $0.43$
\\ \hline
\end{tabular}
\label{tab:results}
\end{center}
\end{table*}

Very recently, the LHCb Collaboration reported their results on possible fully-charm tetraquarks~\cite{Aaij:2020fnh}. They investigated the di-$J/\psi$ invariant mass spectrum, where they observed a broad structure ranging from $6.2$-$6.8$~GeV and a narrow structure at around 6.9~GeV with a global significance of more than $5\sigma$. Their results are in a remarkable coincidence with our previous QCD sum rule predictions~\cite{Chen:2016jxd}, as discussed already in Sec.~\ref{sec:intro}.

Assuming the two structures observed by LHCb~\cite{Aaij:2020fnh} to be compact diquark-antidiquark $[c c] [\bar c \bar c]$ states, in this paper we systematically study their fall-apart decay properties. To do this we use the Fierz rearrangement of the Dirac and color indices to transform diquark-antidiquark $\left([Q Q] [\bar Q \bar Q]\right)$ currents into meson-meson $\left([\bar Q Q] [\bar Q Q]\right)$ currents. The obtained Fierz identities are given in Eqs.~(\ref{eq:tran1}-\ref{eq:tran4}) and Eqs.~(\ref{eq:tran5}-\ref{eq:tran11}), based on which we study their possible fall-apart decays within the naive factorization scheme.

In this paper we have calculated as many as possible relative branching ratios, and the results are given in Eqs.~(\ref{eq:X1}-\ref{eq:X11}) and summarized in Table~\ref{tab:results}. Before drawing conclusions, let us generally discuss about their uncertainty. In the present study we have worked within the naive factorization scheme, so our uncertainty is larger than the well-developed QCD factorization method~\cite{Beneke:1999br,Beneke:2000ry,Beneke:2001ev}, that is at the 5\% level when being applied to study weak and radiative decay properties of conventional (heavy) hadrons~\cite{Li:2020rcg}. On the other hand, the tetraquark decay constants, such as $f_{X_1}$, are removed when calculating relative branching ratios. This significantly reduces our uncertainty. However, it is still not easy to explicitly evaluate our uncertainty, since the corrections to the naive factorization are still not calculable, as of now.

The LHCb experiment~\cite{Aaij:2020fnh} observed a broad structure at around $6.2$-$6.8$~GeV. It can be interpreted as an $S$-wave $cc\bar c \bar c$ tetraquark state, whose mass was predicted to be about 6.5~GeV in our previous QCD sum rule study~\cite{Chen:2016jxd}. To study its fall-apart decays, we use the four $cc\bar c \bar c$ currents, which well correspond to the four $S$-wave $cc\bar c \bar c$ tetraquark states. The obtained results are summarized in the 3rd-6th rows of Table~\ref{tab:results}.

Our results suggest that the broad structure observed by LHCb at around $6.2$-$6.8$~GeV~\cite{Aaij:2020fnh} has the quantum numbers $J^{PC} = 0^{++}$ or $2^{++}$. We propose to confirm it in the di-$\eta_c$ channel. This channel is also helpful to determine its quantum numbers and understand its internal structure. Besides, we propose to search for another $J^{PC} = 1^{+-}$ state in the $J/\psi \eta_c$ channel also at around 6.5~GeV.

The LHCb experiment~\cite{Aaij:2020fnh} observed a narrow structure at around 6.9~GeV. It can be interpreted as a $P$-wave $cc\bar c \bar c$ tetraquark state, whose mass was predicted to be also about 6.9~GeV in our previous QCD sum rule study~\cite{Chen:2016jxd}. To study its fall-apart decays, we investigate all the negative-parity tetraquark currents without derivatives, and the results are summarized in the 7th-13th rows of Table~\ref{tab:results}. Their correspondences to physical states are not so clear. However, if some of them well (better) couples to the physical state, the results extracted from this current should also be (more) consistent with that state. Note that the internal structure of $P$-wave $cc\bar c \bar c$ tetraquark states are still unknown and difficult to be known, since there can be as many as twenty states~\cite{liu:2020eha}.

Our results suggest that the narrow structure observed by LHCb at around 6.9~GeV~\cite{Aaij:2020fnh} has the quantum numbers $J^{PC} = 0^{-+}$ or $1^{-+}$. We propose to confirm it in the di-$\eta_c$, $J/\psi h_c$, $\eta_c \chi_{c0}$, and $\eta_c \chi_{c1}$ channels. These channels are helpful to determine its quantum numbers as well as understand its internal structure. We also propose to search for the $J^{PC} = 0^{--}$ and $1^{--}$ states in the $J/\psi \eta_c$, $J/\psi \chi_{c0}$, $J/\psi \chi_{c1}$, and $\eta_c h_c$ channels at around 6.9~GeV.

To end this paper, we kindly note that there have been many other theoretical studies on the fully-charm tetraquark states, but the obtained conclusions are quite model dependent. Accordingly, there can be many other possible interpretations for the two structures observed by LHCb~\cite{Aaij:2020fnh}, e.g., their possible quantum numbers can be different from those extracted in the present study. Hence, further experimental and theoretical studies are crucial to understand them.

\section*{Conflict of interest}

The authors declare that they have no conflict of interest.

%
\section*{Acknowledgments}
%

This work was supported by
the National Natural Science Foundation of China (11722540 and 11975033),
the Chinese National Youth Thousand Talents Program,
the China National Funds for Distinguished Young Scientists (11825503),
the National Program for Support of Top-notch Young Professionals,
and the Fundamental Research Funds for the Central Universities.

\section*{Author contributions}

All the authors provided the theoretical calculations, discussed the results, and wrote the manuscript.
All the authors have given approval to the final version of the manuscript.

%

%

\end{document}